\documentstyle[11pt,aaspp]{article}
\newcommand{\BE}{\begin{equation}}
\newcommand{\BEAL}{\begin{eqnarray}}
\newcommand{\EE}{\end{equation}}
\newcommand{\EEAL}{\end{eqnarray}}
\def\_#1{_{\scriptscriptstyle #1}}
\def\^#1{^{\scriptscriptstyle #1}}

\def\tot{{2\over 3}}

\def\kms{kms\^{-1}}
\def\cmss{cm~s\^{-2}}
\def\lsun{L\_{\odot}}
\def\msun{M\_{\odot}}
\def\mlsunl{(M/L)\_{\odot}}

\def\ao{a_{o}}
\def\ho{H_0}

\def\slos{\sigma\_{los}}

\def\av#1{\langle #1\rangle}

\begin{document}
\title{Galaxy groups and the modified dynamics}
\author{ Mordehai Milgrom}
\affil{Department of Condensed-Matter
 Physics, Weizmann Institute, Rehovot, Israel}
\begin{abstract}
I estimate Modified-Dynamics (MOND), median $M/L$ values
for recently published catalogues of galaxy groups.
While the median, Newtonian M/L values quoted for these
catalogues are $(110-200)h\_{75}\mlsunl$, the
corresponding values for MOND are less than $10~\mlsunl$ (where the
mass includes contributions from galactic, and intra-group gas).
\end{abstract}
\section{INTRODUCTION}
\par
The modified dynamics (MOND)--propounded as an alternative
to dark matter (\cite{mond})--has been extensively tested on individual
galaxies (e.g. \cite{san}, \cite{mc}, \cite{sanver}, and
 references therein).
The next rung in the ladder--binary galaxies--is notoriously unwieldy
(mainly because of high contamination by false pairs).
The situation improves for small galaxy groups, for example
because identification becomes more secure with increasing number of
members. Still, the uncertainties in the mass determination
 for an individual group are very large, so as to render the result
useless for constraining the amount of dark matter in Newtonian
dynamics, or for testing MOND. This is reflected in the fact that
the dispersion in M/L values deduced for groups is very large.
It is hoped that ``typical'' values deduced for a carefully selected
sample of groups--such as median values for the sample--is a faithful
representation of the dynamics of groups. Newtonian
analyses yield high (median) $M/L$ values for group catalogues:
$M/L\sim (110-200)h\_{75}\mlsunl$.
\par
The only published MOND analysis of groups (\cite{milgr}) was based on
two small group catalogues, and employed a primitive
MOND mass estimator in default of a more adequate one, such as we have
now. Here, I revisit the problem with the much more extensive group
catalogues that have been published in the meanwhile,
using an improved mass estimator.
My main source is \cite{tuck} who list a catalogue gleaned from
the Las Campanas Redshift Survey (LCRS). I also use results based on the
groups identified in the CfA survey: CfA1 (\cite{nw}),
in the Southern-Sky Redshift Survey (SSRS) (\cite{ssrs}), and in the
CfA2 group catalogue (\cite{rpg}), all listed for comparison in
\cite{tuck}. Section 2 describes the estimator used, \S 3 summarizes the
results, and \S 4 discusses the sources of uncertainty.

\section{METHOD}
\par
The MOND mass estimator I use for groups is based on the relation
\BE \av{\av{(v-v_{com})^2}}_t=\tot(MG\ao)^{1/2}[1-\sum_i (m_i/M)^{3/2}]
 \label{nus} \EE
 (\cite{lsy}, \cite{conf}).
Here, $v$ is the 3-D velocity, $v_{com}$ is the center-of-mass velocity,
 $\av{}$  is the mass-weighted average over the constituents, whose
masses are $m_i$, $\av{}_t$ is the
long-time average, and $M$ is the total mass.
 The acceleration constant of MOND is taken to be
$\ao=1.2\cdot 10^{-8}h^2\_{75}\cmss$ (\cite{bbs}).
 Relation(\ref{nus}) is exact
in the deep-MOND limit (all accelerations much smaller than $\ao$)
of the formulation of MOND as modified gravity (\cite{bm}).
Interestingly, the fact that the time-average rms velocity depends
solely on the constituent masses (and not, e.g., on system size)
follows from the conformal invariance of this deep-MOND limit
 (\cite{conf}).
It is also assumed that the system is isolated in the MOND sense;
i.e., is not subject to an external field, and, of course, that it is
decoupled from the Hubble flow (crossing
time is much shorter than the Hubble time).
 To use eq.(\ref{nus}) for deriving total masses from existing
 velocity data we need to make further approximations:
(i) We drop the long-time average, conceding that the estimate might
be in large error for an individual group.
It will still be correct statistically if the sample contains
enough ``replicas'' so that the sample average covers the time average.
(ii) Since only line-of-sight-velocity data is available, I replace
$\av{(v-v_{com})^2}$
 by $3\av{(v-v_{com})^2_{los}}$ (again, assuming that system average
accounts for angular average). Assumptions (i)(ii) are also made
in the Newtonian analysis.
(iii) Group catalogues list, almost exclusively, not the mass-weighted
velocity dispersion that appears in eq.(\ref{nus}),
 but the number-weighted dispersion
 $\slos\equiv[\sum_i(v_i-\hat v)^2_{los}/(N-1)]^{1/2}$, with
$\hat v=\sum_i v_{i,los}/N$, and $N$ the number of member galaxies
in the group.
The luminosity-weighted velocity dispersion would be a better measure
of the mass-weighted one, but the data for individual groups needed to
 calculate it are not available to me, so I shall use $\slos$ instead.
(iv) I approximate the right-hand side of eq.(\ref{nus}) by
$\tot(MG\ao)^{1/2}$, which is valid
 in the limit of a large number of constituents, each having a mass
$\sim M/N\ll M$
 (when all the masses are equal, the correction factor is
 $\sim 1-N^{-1/2}$).
Implementing these approximations we get from eq.(\ref{nus})
\BE M\approx {81\over 4}\slos^4(G\ao)^{-1}. \label{nusa} \EE
This is the group-mass estimator I shall use.
The large majority of the groups in the catalogues I consider
are indeed in the deep-MOND regime (with median acceleration
estimates of only a few percent of $\ao$--see below).
 The effects of the
approximations (iii)(iv) above, as well as a discussion of the effects
of external fields, will be discussed in section 4.

\section{RESULTS}

\par
Table 1 of \cite{tuck} lists the median group properties for their own
sample, together with those for the other catalogues listed above.
The relevant entries (normalized to $\ho=75~\kms~Mpc^{-1}$)
are reproduced in my Table 1 together with the results of the MOND
analysis:
(1) The number of groups in the sample, $N$. (\cite{tuck} have extracted
from their full catalogue of 1495 groups a sub-sample of higher
quality for analysis, containing 394 groups.)
(2) The median line-of-sight velocity dispersion, $\bar\slos$.
(3) The median, Newtonian, dynamical mass, $\bar M_{N}$.
(4) The median, MOND mass, $\bar M_{M}$. This is obtained by substituting
the mean velocity dispersion in eq.(\ref{nusa}).
(5) The median luminosity, $\bar L$ (no luminosity data is given for
the SSRC catalogue).
(6) The median, Newtonian $M/L$ value, $(\overline{M/L})_{N}$.
(7) The Newtonian, median-mass-to-median-luminosity ratio.
(8) The MOND median-mass-to-median-luminosity ratio.
(9) The quantity $\bar a\equiv 3\bar\slos^2/\bar r_h$ (where $\bar r_h$
is the median harmonic group radius)-which is some measure of the typical
acceleration in the groups--in units of $\ao$. We see that indeed
the typical accelerations are much smaller than $\ao$.
\par
 I do not
calculate $M/L$ values for individual groups (the required data is not
even available to me in some cases), and so I do not give median
$M/L$ values for MOND.
One may take $\bar M/\bar L$  as the ``typical'' $M/L$
value for the catalogue.
 This would be somewhat different from the median $M/L$ value,
but the difference is insignificant in comparison with the uncertainties.
For comparison I also give in table 1 the corresponding values for the
Newtonian case.
\par
Note that the catalogue-to-catalogue variations in the MOND
median quantities are larger than those in the corresponding Newtonian
quantities. I believe this is largely due to the high sensitivity
of the MOND mass estimator to the velocity dispersion. A variation
of 50 percent in the median dispersion, as we see here,
is amplified into
 a variation of a factor of 5 in the median masses, for example.
Some inter-catalogue variations are expected because the groups
come from rather different galaxy pools
(e.g. different limiting redshifts), and are selected in different ways.
For example, the LCRS results are base on a small, choice sub-sample
of groups, while no such culling was implemented in the other catalogues.

\par
Considering the uncertainties, one may conclude that the data is
consistent, in the framework of MOND, with no dark matter in galaxy
 groups.

\begin{table}
\caption{Median Newtonian, and MOND, parameters for the group
catalogues}
%\begin{tabular}{|c|cccc|} \hline
\begin{tabular}{ccccc} \hline\hline
    & LCRS & CfA1 & SSRC & CfA2\\
\hline\hline
$N$               & 394 & 166 & 87  & 406\\
$\bar\slos(\kms)$ & 164 & 123 & 183 & 192\\
$\bar M_{N}~(10^{11}h\^{-1}\_{75}\msun)$ & 253 & 207 & 216 & 248\\
$\bar M_{M}~(10^{11}h\^{-2}\_{75}\msun)$ & 9.2 & 2.9 & 14.2 & 17.2\\
$\bar L~(10^{11}h\^{-2}\_{75}\lsun)$ & 2.5 & 1.2 & ... & 2.0 \\
$(\overline{M/L})_{N}~[h\_{75}\mlsunl]$ &  115 & 198 & ... & 180\\
$\bar M_{N}/\bar L~[h_{75}\mlsunl]$ &  102 & 171 & ... & 124\\
$\bar M_{M}/\bar L~\mlsunl$ &  3.7 & 2.4 & ... & 8.6\\
$\bar a/\ao(10^{-2}h\^{-1}\_{75})$ & 2.8 & 1.4 & 4.2 & 5.2 \\
\hline
\end{tabular}
\end{table}

\section{SOURCES OF UNCERTAINTY}
\par
An extensive discussion of
the various sources of possible errors and uncertainties that beset the
selection of groups and their analysis can be found, e.g., in
\cite{rpg}, and in references therein. For example, they estimate (by
comparison with simulations) that some 50\%-75\% of their three-member
groups might be fictitious (10\%-30\% for four-member groups).
Because false groups tend to have higher velocity dispersions this
leads to a systematic overestimate of the masses and M/L values. This
is particularly true in MOND where the mass goes like the fourth power of
the velocity dispersion.
As the surveys are flux limited, the analysis might miss some dimmer
members, so the listed values of the luminosity might be too small.
There is also contributions from gas inside the galaxies (and
possibly from intra-group gas). So the deduced $M/L$ values (Newtonian
and MOND) overestimate stellar $M/L$ values of groups.
\par
There are additional sources of uncertainty that are specific to MOND.
In deriving the mass estimator it was assumed that the system is not
subject to external acceleration fields. When it is, the estimated
MOND masses should be higher. If the external acceleration is comparable
with the intrinsic one, the mass should be increased by about $2^{1/2}$.
It is impracticable to correct the above results for this effect, but we
note that a group will have to be rather near a prominent mass
to have a material effect on its mass estimate.
For example, for a group whose internal acceleration is
 $3\cdot 10^{-2}\ao$--as I find typical--at
 a distance of $20~Mpc$ from the center of a
cluster whose line-of-sight velocity dispersion is $750~\kms$, the effect
is small (the estimated mass has to be increased by about 50\%).
At a distance of $10~Mpc$ the increase is by a factor of about 2.
Nearer yet we have to use a different estimator which is
$M_{M}\sim (a/\ao)M_{N}$. Another example:
for a segment of the Perseus-Pisces large-scale filament \cite{fila}
finds a typical acceleration of $(3-5)10^{-2}\ao$,
 which is similar to what I
find here for groups. So, for groups in an environment such as this,
$M$ will have to be corrected up by $\sim 2^{1/2}$.
\par
To get an idea of the error introduced by making approximations (iii)
and (iv) in the mass estimator, I now consider
 two special cases of stationary, and isotropic groups
for which the comparison
is simple: (a) a group made of $N$ equal masses $m$. (b) A group
comprising one massive galaxy with all the others of
 negligible (and equal) test masses.
In case (a) the ratio $\eta$ of the correct mass to that given by
 eq.(\ref{nusa}) is
\BE \eta=\left({1-N^{-1} \over 1- N^{-1/2}}\right)^2.\label{cora} \EE
So, eq.(\ref{nusa}) {\it underestimates} the mass by a factor of 2.5 for
$N=3$, a factor 9/4 for $N=4$, and so on. Case (b): we use the fact
that for a test particle in an arbitrary
(low acceleration) orbit around a mass $M$, the line-of-sight velocity,
averaged over the orbit and over all lines of sight is given by
$\av{v^2}=MG\ao/3$; hence, for a system as in (b)
$M\approx 9\slos^4(G\ao)^{-1}$.
 The correction factor is thus
 4/9, and eq.(\ref{nusa}) {\it overestimates} the mass by a
factor of 9/4.
\par
The uncertainties introduced by these last two MOND-related
 approximations are, by and large, small compared with those
associated with group identification and unknown
 geometry, and do not change
the conclusion that group dynamics is consistent, within the framework
of MOND, with no dark matter in groups.

\acknowledgements
I thanks the referee, Marc Verheijen, for useful comments.

%\clearpage

\end{document}